\documentclass[twocolumn,prd,superscriptaddress,showpacs,preprintnumbers,amsmath,amssymb,nofootinbib]{revtex4}

\usepackage{subfigure}
\usepackage{amsmath,amssymb}
\usepackage{graphicx}
\usepackage{rotating}
\usepackage{bm}
\usepackage{dsfont}
\usepackage{color}

\newcommand{\smallstep}{\vspace{.0em}}
\def\di{\displaystyle}

\def\bg{\begin{eqnarray}\begin{array}{rcl}\displaystyle}
\def\eg{\end{array} &\di    &\di   \end{eqnarray}}
\def\bm#1{\begin{eqnarray}\begin{array}{#1}\di}
\def\bmo#1{\begin{eqnarray*}\begin{array}{#1}\di}
\def\bml#1#2{\begin{eqnarray}\begin{array}{#1}\label{#2}\di}
\def\bgo{\begin{eqnarray*}\begin{array}{rcl}\displaystyle}
\def\ego{\end{array} &\di    &\di \nonumber  \end{eqnarray*}}

\def\btensor#1#2{\renew\left#1\begin{array}{#2}\di}
\def\brtensor#1#2#3{\ren#3\left#1\begin{array}{#2}}
\def\botensor#1#2{\renew\left#1\begin{array}{#2}}
\def\etensor#1{\end{array}\right#1}

\def\eq#1{(\ref{#1})}
\def\Eq#1{Eq.~(\ref{#1})}

\def\s0#1#2{\mbox{\small{$ \frac{#1}{#2} $}}}
\def\0#1#2{\frac{#1}{#2}}

\def\CP{{\mathcal P}}

\def\Z{\mathds{Z}}
\def\ren#1{\renewcommand{\arraystretch}{#1}}

\def\renew{\renewcommand{\arraystretch}{1}}
\definecolor{blue}{rgb}{0,0,1}

\definecolor{green}{rgb}{0,1,0}

\definecolor{red}{rgb}{1,0,0}

\newcommand{\Tr}{\mathrm{Tr}}

\newcommand{\tr}{\mathrm{tr}}

\newcommand{\I}{\mathrm{i}}
\newcommand{\be}{\begin{eqnarray}}
\newcommand{\ee}{\end{eqnarray}}

\newcommand{\Nc}{N_{\text{c}}}

\begin{document}

\title{Confinement in Polyakov Gauge}

%\preprint{HD-THEP-09-??}

%05.10.Cc Renormalization group methods
%12.38.Aw General properties of QCD (dynamics, confinement, etc.)
%11.10.Wx Finite-temperature field theory

\author{Florian Marhauser}

\author{Jan M.~Pawlowski}

\affiliation{Institut f\"ur Theoretische Physik, Universit\"at Heidelberg, 
Philosophenweg 16,
D-69120 Heidelberg, Germany}

\begin{abstract}
%\abstract{
  We approach the non-perturbative regime in finite temperature QCD
  within a formulation in Polyakov gauge. The construction is based on
  a complete gauge fixing. Correlation functions are then computed
  from Wilsonian renormalisation group flows.  First results for the
  confinement-deconfinement phase transition for $SU(2)$ are
  presented. Within a simple approximation we obtain a second order
  phase transition within the Ising universality class. The critical
  temperature is computed as $T_c \simeq 305$ MeV. 
%}
\end{abstract}

\maketitle

\section{Introduction}\label{sec:intro}

One of the remaining problems in low energy QCD is the quantitative 
field theoretical description of the confinement-deconfinement phase 
transition. Apart from its genuine importance for a first principle
understanding of the confining physics in QCD it also is a key input
for the evaluation of the QCD phase diagram. 

In the past decade much progress has been achieved both in continuum
studies as well as with lattice computations for our understanding of
the low energy sector of QCD, for reviews see e.g.\
\cite{Svetitsky:1985ye,Alkofer:2000wg,Litim:1998nf,Schaefer:2006sr,%
Fischer:2008uz}.
For an analytical description of the low energy sector, topological
degrees of freedom are likely to play an important role for the
confinement-deconfinement phase transition as well as for chiral
symmetry breaking, see e.g.\ \cite{Schafer:1996wv}. The latter has
been very successfully described within instanton models, whereas the
confining properties of the theory are harder to incorporate within
semi-classical descriptions.  Indeed, tracking down those topological
degrees of freedoms relevant for confinement in the physical vacuum
has its intricacies as the physical vacuum is more likely to contain a
rather dense packing of topological configurations, making their
detection difficult.  Moreover, models of confinement are rather based
on topological defects instead of stable topological objects, the
construction of which out of these defects is plagued by
non-localities.\smallstep

Still, these defects are manifest in the Polyakov loop, the order
parameter in pure Yang-Mills theory \cite{Polyakov:1978vu}, and can be
extracted by an appropriate gauge fixing, see e.g.\
\cite{Reinhardt:1997rm,Ford:1998bt}. Gauge fixing is also
mandatory in most continuum formulations of QCD for removing the
redundant gauge degrees of freedom.  This is mostly seen as a
liability of such an approach, as a formulation of QCD in
gauge-variant variables complicates the access to gauge invariant
observables. However, gauge fixing is nothing but a reparameterisation
of the path integral and can be used for even facilitating the
computation of at least a subset of observables.  Indeed, this point
of view has been exploited much in the discussion of confinement
mechanisms based on topological defects.  More recently it also has
become clear that these are not competing physics mechanisms but
rather different facets of the same global physics picture which still
awaits a fully gauge invariant description, see e.g.\
\cite{Greensite:2004mh}. Despite this final step we have learned much
from the combined investigations which together built a nearly
complete mosaic.  \smallstep

The effective potential of the Polyakov loop has also been used as an
input for effective field theories that give some access to the QCD
phase diagram \cite{Pisarski:2000eq}.  At finite temperature and
vanishing density, these models have led to impressive results in
particular for thermodynamical quantities. At finite chemical
potential, the back-reaction of the matter sector to the gauge sector
is difficult to quantify in these models, and the chiral and
confinement-deconfinement phase transitions are sensitive to the
details of this back-reaction. This also holds true for the question
of a quarkyonic phase with confinement and chiral symmetry at finite
density \cite{McLerran:2007qj}. For an extension of these models one
has to resort to a field-theoretical description of the gauge sector
which allows to systematically study the impact of a finite chemical
potential on the dynamics of the gauge field,
\cite{Braun:2008pi}.\smallstep

In summary, the evaluation of Green functions of the Polyakov loop
allows for a direct access to the physics in the strongly coupled
sector of QCD, and in particular the confinement-deconfinement phase
transition. In the present work we initiate a non-perturbative study
of QCD in Polyakov gauge. In this gauge the Polyakov loop takes a
particularly simple form and is directly related to the temporal
component of the gauge field. After integrating-out the spatial
components of the gauge field, and formulated with Polyakov loop
variables, the gauge field sector of QCD resembles a scalar model. The
dynamics of low energy Yang-Mills theory is then incorporated by
evaluating Wilsonian flows for the effective action
\cite{Wetterich:1993yh,Litim:1998nf,Schaefer:2006sr,Berges:2000ew,%
  Bagnuls:2000ae,Pawlowski:2005xe}. We derive the flow equation for
QCD in Polyakov gauge, and solve it for the full effective potential
of the Polyakov loop. Due to the formulation in Polyakov gauge a
simple truncation already suffices to encode the physics of the
confinement-deconfinement phase transition. The results include the
temperature dependence of the Polyakov loop, and the critical
temperature. We also compare the present approach to lattice studies
\cite{Fingberg:1992ju}, and to a recent continuum computation in
Landau gauge \cite{Braun:2007bx}.

\section{QCD in Polyakov gauge}\label{sec:QCDinPol}

In QCD with static quarks the expectation value of a static quark
$\langle q(\vec x)\rangle$ serves as an order parameter for
confinement. It is proportional to the free energy $F_q$ of such a
state, $\langle q(\vec x)\rangle\sim \exp(-\beta F_q)$, where $\beta
=1/T$ is the inverse temperature. Hence in the confining phase at low
temperature, the expectation value vanishes, whereas at high
temperatures in the deconfined phase, it is non-zero.  The Polyakov
loop variable, \cite{Polyakov:1978vu}, is the creation operator for a
static quark,
\begin{equation}\label{eq:Polloop}
L(\vec x)=\frac{1}{\Nc} \tr\, \CP(\vec x)\,, 
\end{equation}
where the trace in \eq{eq:Polloop} is done in the fundamental
representation, and the Polyakov loop operator is a Wegner-Wilson 
loop in time direction, 
\begin{equation}\label{eq:Polop}
{\cal P}(\vec x)  =\CP\, \exp \left( \I g
  \int_0^\beta dx_0\, {A}_0(x_0,\vec x) \right)\,. 
\end{equation} 
Here ${\cal P}$ stands for path ordering. We conclude that $\langle
q(\vec x)\rangle \simeq \langle L(\vec x)\rangle$, and thus the
negative logarithm of the Polyakov loop expectation value relates to
the free energy of a static fundamental color source. Moreover,
$\langle L\rangle$ measures whether center symmetry is realised by the
ensemble under consideration, see e.g.\
\cite{Polyakov:1978vu,Svetitsky:1985ye,Schafer:1996wv,%
Reinhardt:1997rm,Ford:1998bt,Greensite:2004mh}.

More specifically we consider gauge transformations $U(x_0,x)$ with
$U(0,\vec x) U^{-1}(\beta, \vec x) =Z$, where $Z$ is a center element.
In $SU(2)$ the center is $Z_2$, whereas in physical QCD with $SU(3)$ 
it is $Z_3$. Under such center transformations the Polyakov loop
operator ${\cal P}(\vec x)$ in \eq{eq:Polop} is multiplied with 
a center element $Z$, 
\begin{equation} \label{eq:centertrafo}
{\cal P}(\vec x)\to Z\,{\cal P}(\vec x)\,,
\end{equation} 
and so does the Polyakov loop, $L(\vec x)\to Z\, L(\vec x)$.  Hence, a
center-symmetric confining (disordered) ground state ensures $\langle
L\rangle=0$, whereas deconfinement $\langle L\rangle\neq 0$ signals
the ordered phase and center-symmetry breaking,
\begin{eqnarray}\nonumber  
T<T_c: &\qquad \langle L(\vec x)\rangle = 0\,,\quad F_q=\infty\,, \\ 
 T>T_c: &\qquad \langle L(\vec x)\rangle \neq 0\,, \quad F_q<\infty\,.
\label{eq:orderdisorder}\end{eqnarray}
The expectation value of the Polyakov loop can be deduced from the
equations of motion of its effective potential $V_L[\langle
L\rangle]$. We shall argue, that the computation of the latter greatly
simplifies within an appropriate choice of gauge. Indeed, gauge fixing
is nothing but the choice of a specific parameterisation of the path
integral, and a conveniently chosen parameterisation can simplify the
task of computing physical observables. 

In the present case our choice of gauge is guided by the demand of a
particularly simple representation of the Polyakov loop variable
\eq{eq:Polloop}. A gauge ensuring time-independent $A_0$ leads to
both, a trivial integration in \eq{eq:Polop} and renders the path
ordering irrelevant. Having done this one can still
rotate the Polyakov loop operator ${\cal P}(\vec x)$, \eq{eq:Polop},
into the Cartan subgroup. The above properties are achieved for
time-independent gauge field configurations in the Cartan subalgebra,
i.e.\ $A_0(t_0,\vec x)=A_0^c(\vec x)$. For $SU(2)$ this reads 
\begin{eqnarray}\label{eq:A0}
A_0(t_0,\vec x)=A_0(\vec x)\, \0{\sigma_3}{2}\end{eqnarray} 
and entails a
particularly simple relation between $A_0$ and $L$,
\begin{equation} \label{eq:polsu(2)}
L(\vec x) = \cos\, \012 g \beta A_0(\vec x) \,, 
\end{equation} 
Note that this simple relation is not valid on the level of
expectation values of $L$ and $A_0$, in $SU(2)$ we have $\langle
L\rangle \neq \cos\, \012 g \beta \langle A_0\rangle$. However, in the
present work we consider an approach that gives direct access to the
effective potential $V_{\rm eff}[ \langle A_0\rangle]$ for the gauge
field, as distinguished to those for the Polyakov loop, $U_{\rm
  eff}[\langle L\rangle]$ \footnote{A reformulation in terms of the
  Polyakov loop variable only along the lines outlined in
  \cite{Pawlowski:2005xe} is in progress.}.

Here, we argue that $ L[\langle A_0\rangle ]$ also serves as an order
parameter: to that end we show that the order parameter $ \langle L[
A_0 ]\rangle $ is bounded from above by $ L[\langle A_0\rangle ]$.  It
follows that $ L[\langle A_0\rangle ]$ is non-vanishing in the
center-broken phase. Furthermore we show that in the center-symmetric
phase with vanishing order parameter, $\langle L[ {A_0}] \rangle=0$, 
also the observable $L[\langle {A_0} \rangle]$ vanishes. For the sake
of simplicity we restrict the explicit argument to $SU(2)$, but it
straightforwardly extends to general $SU(N)$. First we note that we
can use \eq{eq:polsu(2)} for expressing the expectation value of $A_0$
in terms of $L$,
\begin{equation} \label{eq:A_0fromL}
\012 g \beta  \langle A_0\rangle =\langle \arccos L\rangle\,.
\end{equation}
We emphasise that the rhs of \eq{eq:A_0fromL} defines an observable as
it is the expectation value of an gauge invariant object. This
observable happens to agree with $\langle A_0\rangle $ in Polyakov
gauge. It follows from the Jensen inequality that the expectation
value of the Polyakov loop, the order parameter for confinement, is
bounded from above by $L[\langle {A_0} \rangle]$, see
\cite{Braun:2007bx}
\begin{eqnarray}\label{eq:jensen} 
L[\langle {A_0} \rangle]\geq \langle L[A_0]\rangle\,. 
\end{eqnarray} 
for gauge fields $g\beta \langle A_0\rangle/2 \in [0,\pi/2]$.  Note
that it is sufficient to consider the above interval due to
periodicity and center symmetry of the potential. This means we
restrict the Polyakov loop expectation value to $\langle L\rangle \geq
0$. Negative values for $\langle L\rangle$ are then obtained by center
transformations, $L\to \pm L$. \Eq{eq:jensen} is easily proven for
$SU(2)$ from \eq{eq:polsu(2)} as $\cos(x)$ is concave for $x\in
[0,\pi/2]$. Thus, for $\langle L\rangle >0$ it necessarily also
follows that $g\beta \langle A_0\rangle/2 <\pi/2$.

In turn we can show that $g\beta \langle A_0\rangle/2=\pi/2$, if the
Polyakov loop variable $\langle L[A_0]\rangle$ vanishes. This then
entails that $L[\langle {A_0} \rangle]=0$. To that end we expand $L$
about its mean value $\langle L\rangle $, that is $L=\langle L\rangle
+\delta L$. Inserting this expansion into \eq{eq:A_0fromL} we arrive
at
\begin{equation} \label{eq:expandA_0} \012 g \beta \langle A_0\rangle
  =\arccos \langle L\rangle-\0{1}{\sqrt{1- \langle L\rangle ^2} } 
\langle \delta L\rangle +
  O\left(\langle \delta L^2\rangle\right)\,.
\end{equation}
In the center-symmetric phase $\langle L\rangle =0$, c.f.
\eq{eq:orderdisorder}.  Under center transformations $L$ transforms
according to (\ref{eq:centertrafo}) $L\to Z\, L$ with $Z=\pm 1$ and
hence $ \delta L\to Z\, \delta L$. It follows that $\langle \delta
L^{2n+1}\rangle = Z\langle \delta L^{2n+1}\rangle =0$, and all odd
powers in \eq{eq:expandA_0} vanish. The even powers vanish since
$\arccos$ is an odd function and hence has vanishing even Taylor
coefficients $\arccos^{(2n)}(0)$. Thus, in the center-symmetric phase
we have
\begin{equation} \label{eq:expandA_0centersym} \012 g \beta \langle
  A_0\rangle =\arccos \langle L\rangle=\0{\pi}{2}\,.
\end{equation}
In summary we have shown 
\begin{eqnarray}\nonumber  
  T<T_c: &\di\quad  L[\langle {A_0} \rangle]=0
  \quad \Leftrightarrow \quad\012 g \beta  \langle A_0(\vec x)\rangle
  = \0\pi2\,,
  \\ 
  T>T_c: &\di\quad L[\langle {A_0} \rangle]\neq 0 \quad \Leftrightarrow
  \quad 
  \012 g \beta  \langle A_0(\vec x)\rangle 
  <\0\pi2\,.
\label{eq:A0orderdisorder}\end{eqnarray}
We conclude that $\langle A_0\rangle $ in Polyakov gauge serves as an
order parameter for the confinement-deconfinement (order-disorder)
phase transition, as does $L[\langle A_0\rangle]$. Thus, we only have
to compute the effective potential $V_{\rm eff}[\langle A_0\rangle]$
in order to extract the critical temperature, and e.g.\ critical
exponents. This potential is more easily accessed than that for the
Polyakov loop. It is here were the specific gauge comes to our aid as
it allows the direct physical interpretation of a component of the
gauge field. This property has been already exploited in the
literature, where it has been shown that $\langle{A}_0\rangle$ in
Polyakov gauge is sensitive to topological defects related to the
confinement mechanism \cite{Reinhardt:1997rm,Ford:1998bt}.

\section{Quantisation}\label{sec:quant}

We proceed by discussing the generating functional of Polyakov gauge 
Yang-Mills theory. For its derivation we use the Faddeev-Popov method. 
Specifying to $SU(2)$, the Polyakov gauge \eq{eq:A0}  
is implemented by the gauge fixing conditions
\begin{equation}\label{eq:Pol1}
 \partial_0 \tr\,\sigma_3 A_0=0\,,  \qquad 
  \tr\,(\sigma_1\pm i\sigma_2) A_0 = 0\,,
\end{equation} 
where the $\sigma_i$ are the Pauli matrices.  However, the gauge
fixing \eq{eq:Pol1} is not complete. It is unchanged under
time-independent gauge transformations in the Cartan sub-group. These
remaining gauge degrees of freedom are completely fixed by the
following conditions,
\begin{eqnarray}\nonumber 
& \partial_1 \int dx_0  \, \tr\, \sigma_3 A_1 =0\,,\qquad  
 \partial_2 \int dx_0 dx_1  \, \tr \, \sigma_3 A_2 = 0\,,&\\ 
& \partial_3  \int dx_0 dx_1 dx_2\,  \tr \,\sigma_3 A_3 = 0\,.&  
\label{eq:Pol2}\end{eqnarray}  
The gauge fixings \eq{eq:Pol2} are integral conditions and are the
weaker the more integrals are involved. Basically they eliminate the
corresponding zero modes. This can be seen directly upon putting the
theory into a box with periodic boundary conditions, $T^4$, see e.g.\
\cite{Ford:1998bt}.
 
The gauge fixing conditions \eq{eq:Pol1},\eq{eq:Pol2} lead to the 
Faddeev-Popov determinant 
\begin{eqnarray}
\Delta_{FP}[A] = (2 T)^2 \left[ \prod_{x} \sin^2 \left( \frac{g A_0^3
      (\vec x)}{ 2 T } \right) \right]\,, 
\label{eq:FPdet}\end{eqnarray}
the computation of which is detailed in appendix~\ref{app:FPdet}.  The
integration over the longitudinal gauge fields precisely cancels the
Faddeev-Popov determinant in the static approximation $\partial_i
A_0^c=0$, see Appendix~\ref{app:FPdet}. Finally we are left with the
action
\begin{eqnarray}  
S_{\rm eff}[A] &\simeq & -\frac{1}{2} \beta \int d^3x\, A_0 
\vec \partial{\,}^2 A_0\\\nonumber 
&&\hspace{-1cm}-\frac{1}{2} \int_T d^4x\, A^a_{\bot,i} \left[(D_0^2)^{ab} +
\vec \partial^2\delta^{ab} \right] A_{\bot,i}^a +O(A_{\bot,i}^3)
\end{eqnarray} 
with $D_0^{ab} = \partial_0 \delta^{ab} + A^3_0 g f^{a3b}$ and
transversal spatial gauge fields, $\partial_i A_{\bot,i}=0$. The
generating functional of Yang-Mills theory in Polyakov gauge then
reads
\begin{eqnarray}\nonumber 
\hspace{-.5cm}Z[J]&=&\int dA_0\,dA_{\bot,i} \,\exp\Bigl\{-S_{\rm eff}[A]\\
& &\hspace{.3cm}
 +\int d^3 x\,J_0 A_0+\int_T d^4 x \,
J_{\bot,i} A_{\bot,i} \Bigr\}\,.
 \label{eq:Zpol}\end{eqnarray}
In \eq{eq:Zpol} we have normalised the temporal component $J_0$ of the
current with a factor $\beta$. The classical action $S_{\rm eff}$ is
inherently non-local as is contains one-loop terms, the Faddeev-Popov
determinant as well as the integration over the longitudinal gauge fields. 

Instead of computing $Z[J]$ in \eq{eq:Zpol} we shall compute the
effective action $\Gamma$ within a functional renormalisation group
approach
\cite{Wetterich:1993yh,Litim:1998nf,Schaefer:2006sr,Berges:2000ew,%
Bagnuls:2000ae,Pawlowski:2005xe}.
To that end we introduce an infrared cut-off for the transversal
spatial gauge fields and in the temporal gauge fields by modifying the
action, $S\to S_{\rm eff}+\Delta S_k[A_0]+\Delta S_{\bot,k}[\vec
A_{\bot}]$, with infrared scale $k$, and cut-off terms
\begin{eqnarray}\nonumber 
  \Delta S_k[A_0]&=&\012 \beta \int d^3 x \,
  A_0\, R_{0,k}\, A_0\\ 
\Delta S_{k,\bot}[\vec A_{\bot}]&=&
 \int_T d^4 x\, A^a_{\bot, i}\, 
  R_{\bot,k}\, A^a_{\bot, i}\,.
\label{eq:Cutoff} \end{eqnarray}
The regulators $R_k$ in \eq{eq:Cutoff} are chosen to be
momentum-dependent and required to provide masses at low momenta
and to vanish at large momenta. For $k\to 0$ they vanish identically.

They can be written as one single regulator $R_{A,{\mu\nu}}$, 
which is a block-diagonal matrix in field space
with entries $R_{A,{00}}=R_{0,k}$ and $R_{A,{ij}}=R_{\bot,k}
\Pi_{\bot,ij}$, where the transversal projector is defined by
\begin{eqnarray}\label{eq:transverse}
\Pi_{\bot,ij} = \delta_{ij} -\frac{p_i p_j}{\vec p^2}\,.
\end{eqnarray} 
The above structure is induced by the fact the $A_{\bot,i}$ are
transversal, and hence $R_{\bot,k}$ only couples to the transversal
degrees of freedom.

The flow of the cut-off dependent effective action $\Gamma_k$ is then
given by Wetterich's equation
\cite{Wetterich:1993yh,Berges:2000ew,Bagnuls:2000ae} for Yang-Mills
theory \cite{Litim:1998nf,Pawlowski:2005xe} in Polyakov gauge,
\begin{eqnarray}\nonumber 
\hspace{-.5cm}  \partial_t \Gamma_{k}& =
  &  \frac{\beta}{2} 
  \int \0{d^3 p}{(2 \pi)^3} \left(\frac{1}{\Gamma_k^{(2)} 
    + R_A}\right)_{00}\partial_t R_{0,k}\\ 
  & & + 
  \frac{T}{2} \sum_{n\in \Z}
  \int \0{d^3 p}{(2 \pi)^3} \left(\frac{1}{\Gamma_k^{(2)} 
    + R_A}\right)_{ii}\partial_t R_{\bot, k}\,,
\label{eq:flow}\end{eqnarray}
where $t$ is the RG time $t = \ln (k / \Lambda)$, and $\Lambda$ is
some reference scale. 

\section{Approximation scheme} \label{sec:approx}

\Eq{eq:flow} together with an initial effective action at some initial
ultraviolet scale $k=\Lambda_{\rm UV}$ provides a definition of the
full effective action at vanishing cut-off scale $k=0$ via the
integrated flow. For the solution of \eq{eq:flow} we have to resort to
approximations to the full effective action. In gauge theories such an
approximation also requires the control of gauge invariance, see e.g.\ 
\cite{Pawlowski:2005xe}. 

Here we shall argue that in Polyakov gauge a rather simple
approximation to the full effective action already suffices to
describe the confinement-deconfinement phase transition, and, in
particular, to estimate the critical temperature. We compute the flow
of the effective action $\Gamma[A_0,\vec A_{\bot}]$ in the following
truncation
\begin{eqnarray}\nonumber  
  \hspace{-.4cm}\Gamma_k[A_0,\vec A_{\bot}] &\!=&\! \beta 
\int d^3x\, \left( 
    -\0{Z_{0}}{2} A_0 
\vec \partial{\,}^2 A_0+V_{k}[A_0]\right) \\
&&\hspace{-.8cm}-\frac{1}{2} \int_T d^4x\, Z_{i} \vec 
A_{\bot}^a \left[(D_0^2)^{ab} +
\vec \partial{\,}^2\delta^{ab} \right] \vec A_{\bot}^a\,,  
\label{eq:effact}\end{eqnarray} 
with $k$-dependent wave function renormalisations $Z_0,Z_i$. 
The effective action \eq{eq:effact} relates to the order parameter
$\langle L(\vec x)\rangle$ as well as its two point correlation
$\langle L(\vec x) L^\dagger (\vec y) \rangle$ via the effective
potential $V_{\rm eff}[A_0]=V_k[A_0]$ as explained in
section~\ref{sec:QCDinPol}. The expectation value $\langle L(\vec
x)\rangle$, or $L[\langle A_0\rangle]$, is used to
determine the phase transition temperature $T_c$ as well as critical
exponents. The temperature-dependence of the Polyakov loop two-point
function relates to the string tension. In the confining phase, for
$T<T_c$, and large separations $|\vec x-\vec y|\to\infty$, the
two-point function falls off like
\begin{equation}\label{eq:string} 
  \lim_{|\vec x-\vec y|\to\infty} 
\langle L(\vec x) L^\dagger (\vec y) \rangle_{c} \simeq \exp 
\left\{-\beta\, \sigma |\vec x-\vec y|\right\}\,. 
\end{equation} 
Here, $\langle\cdots \rangle_c$ stands for the connected part of the 
related correlation function, i.e.\ $ 
\langle L(\vec x) L^\dagger (\vec y) \rangle_{c} =
\langle L(\vec x) L^\dagger (\vec y) \rangle-\langle L(\vec x)\rangle 
\langle L(\vec y)\rangle$. 
In turn, its Fourier transform shows the momentum dependence 
\begin{equation}\label{eq:stringmomentum} 
  \lim_{|p|\to 0} 
  \langle L(0) L^\dagger (p) \rangle_c \simeq  \lim_{|p|\to 0} 
\0{1}{\pi^2} \0{\beta\sigma} {
((\beta\sigma)^2+p^2)^2}=\0{1}{\pi^2} \0{1} {
(\beta\sigma)^3} \,. 
\end{equation} 
We conclude that the Polyakov loop variable has a massive propagator.
This directly relates to a massive propagator of $A_0$ in Polyakov
gauge. 

The approximation scheme is fully set by specifying the regulators
$R_{0,k}$ and $R_{\bot,k}$. Naively one would identify the cut-off
parameter $k$ in the regulators with the physical cut-off scale
$k_{\rm phys}$. For general regulators this is not possible and one
deals with two distinct physical cut-off scales, $k_{0,\rm phys}$ and
$k_{\bot,\rm phys}$ related to $R_{0,k}$ and $R_{\bot,k}$
respectively, for a detailed discussion see \cite{Pawlowski:2005xe}.
However, within the approximation \eq{eq:effact} it is crucial to have
a unique effective cut-off scale $k_{\rm phys}=k_{0,\rm
  phys}=k_{\bot,\rm phys}$, as different physical cut-off scales
$k_{0,\rm phys}\neq k_{\bot,\rm phys}$ necessarily introduce a
momentum transfer into the flow which carries part of the physics.
This momentum transfer is only fully captured with a non-local
approximation to the effective action. In turn, a local approximation
such as \eq{eq:effact} requires $k_{0,\rm phys}=k_{\bot,\rm phys}$. In
other word, a local approximation works best if the momentum transfer
in the flow is minimised. More details about such a scale matching and
its connection to optimisation \cite{Litim:2000ci,Pawlowski:2005xe}
can be found in \cite{Pawlowski:2005xe}. Note in this context that in
the present case we also have to deal with the subtlety that $A_0$
only depends on spatial coordinates whereas $\vec A_\bot$ is
space-time dependent.  However, the requirement of minimal momentum
transfer in the flow is a simple criterion which is technically
accessible.

More specifically we restrict ourselves to regulators \cite{Litim:2006ag} 
\begin{equation} 
R_{A,00} = Z_0 R_{{\rm opt},k}(\vec p^2)\,,\quad 
R_{A,ij} = Z_i \Pi_{\bot,ij}(\vec p)R_{\rm opt,k_\bot }(\vec p^2)\,, 
\label{eq:cutoffs}\end{equation} 
where \cite{Litim:2000ci}
 \begin{eqnarray} 
R_{{\rm opt},k}(\vec p^2)=(k^2-\vec p^2)\theta(k^2-\vec p^2)\,. 
\label{eq:opt}\end{eqnarray} 
The detailed scale-matching argument is deferred to
Appendix~\ref{app:match}, and results in a relation
$k_{\bot}=k_\bot(k)$ depicted in Fig.~\ref{fig:kbotk} in the appendix.
It is left to determine the effective cut-off scale $k_{\rm phys}$.
This cut-off scale can be determined from the numerical comparison of
the flows of appropriate observables: one computes the flow with the
three-dimensional regulator $R_{{\rm opt},k_\bot }(\vec p^2)$ in
\eq{eq:cutoffs}, as well as with the four-dimensional regulator
$R_{{\rm opt},k_{\rm phys}}(p^2)$. Then the respective physical scales
are identified, i.e.\ $k_{\bot,\rm phys}(k_\bot)=k_{\rm phys}$. The
results for this matching procedure are depicted in
Fig.~\ref{fig:kskphys} in Appendix~\ref{app:match}.  Another estimate
comes from the flow related to the three-dimensional
$A_0$-fluctuations, where we can directly identity $k_{\rm phys}=k$.
We use the above choices as limiting cases for an estimate of the
systematic error in our computation. Our explicit results are obtained
for the best choice that works in all physics constraints.

\section{Flow}\label{sec:flow}

\begin{figure}[h]\vspace{.3cm}
  \includegraphics[width=8cm]{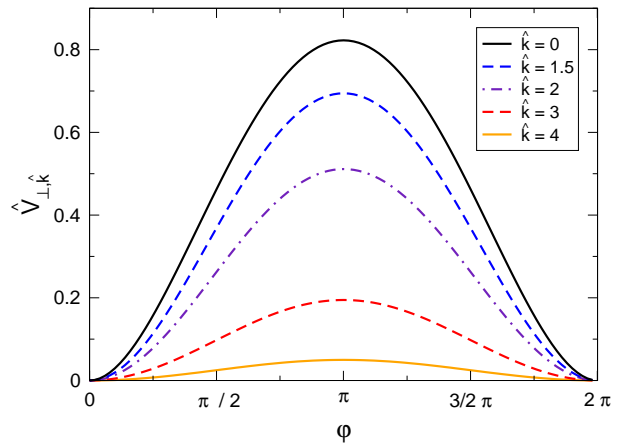}\\
  \caption{$\hat V_{\bot}$ for different values of $\hat k$}
  \label{fig:VPreWeiss3D}
\end{figure}

We are now in the position to integrate the flow equation
\eq{eq:flow}.  To begin with, we can immediately integrate out the
spatial gauge fields $\vec A_\bot$ for $Z_i=1$, that is the second
line in \eq{eq:flow}. This part of the flow only carries an explicit
dependence on the cut-off $k$, details of the calculation can be found
in Appendix \ref{app:intoutAI}. It results in a non-trivial effective
potential $V_{\bot,k}[A_0]$ that approaches the Weiss potential
\cite{Weiss:1980rj} in the limit $k/T \to 0$, and falls off like
$\exp(-\beta k_\bot (k)) \cos (g \beta A_0) $ in the UV limit $k/T\to
\infty$, see Fig.~\ref{fig:VPreWeiss3D}. In terms of the effective
action, after the integration over $\vec A_\bot$, we are led to an
effective action of $A_0$,
\begin{equation}
  \label{eq:truncated_eff_actioncopy}
\Gamma_{k}[A_0] = \beta \int d^3x \left(\frac{ Z_0}{2} 
(\vec \partial A_0)^2 + \Delta V_{k}[A_0] + V_{\bot,k} [A_0] \right)\,. 
\end{equation}
\Eq{eq:truncated_eff_actioncopy} follows from \eq{eq:effact} with 
$\Gamma_k[A_0]=\Gamma_k[A_0,\vec A_\bot =0]$, and 
\begin{equation}
  \label{eq:Vk}
V_k[A_0]= \Delta
V_{k}[A_0] + V_{\bot,k} [A_0]\,. 
\end{equation}
The full effective potential is given by $V_{\mathrm{eff}}[A_0] =
\Delta V_{k=0}[A_0] + V_{\bot,k=0}[A_0]$. We are left with the task to
determine $\Delta V_k$, which is the part of the effective potential
induced by $A_0$-fluctuations. In Polyakov gauge, these fluctuations
carry the confining properties of the Polyakov loop variable, whereas
the spatial fluctuations generate a deconfining effective potential
for $A_0$, see Appendix~\ref{app:intoutAI}. We emphasise that this
structure is not present for spatial confinement which is necessarily
also driven by the spatial fluctuations, and solely depends on these
fluctuations in the high temperature limit. We hope to report on this
matter in the near future.

Here we proceed with the integration of the flow for the potential
$\Delta V_{k}$. To that end we reformulate the flow \eq{eq:flow} as a
flow for $\Delta V_{k}$ with the external input $V_{\bot,k}$, see
\eq{eq:flowDeltaVapp}. The flow equation for $\Delta V_{k}$ reads
\begin{equation} 
\label{eq:deltaV_FRGeq} 
\beta\, \partial_{t} {\Delta V}_{k} =
\frac{1}{2} \int \0{d^3 p}{(2 \pi)^3} \frac{ \partial_{t}
    R_{0,k}}{Z_0\vec p{\,}^2 +
\partial^2_{A_0}( \Delta V_{k} + V_{\bot,k}) +
    R_{0,k}} \,.  
\end{equation}
With the specific regulator $R_k$ in \eq{eq:cutoffs} we can perform
the momentum integration analytically. We also introduce the scalar
field $\varphi = g \beta A_0$, and arrive at 
\begin{eqnarray}\label{eq:preflowV}
  \beta \partial_k \Delta V_k
  = \frac{2}{3 (2 \pi)^2} \frac{(1+\eta_0/5) k^2 }{1+\frac{ g_{k}^2 
\beta^2}{ k^2 }
      \partial^2_{\varphi} ( V_{\bot,k} + \Delta V_k)}\,, 
\end{eqnarray}
where the coupling $g_k^2$ has to run with the effective cut-off
scale $k_{\rm phys}$, and is estimated by an appropriate 
choice of the running coupling $\alpha_s$, 
\begin{equation}\label{eq:runningg}
g_k^2=\frac{g^2}{Z_0}\,,\qquad {\rm with}\qquad g_k^2=4 \pi 
\alpha_s(k_{\rm phys}^2)\,,
\end{equation} 
see also Appendix~\ref{app:intoutAI}. This also entails that the anomalous 
dimension $\eta_0$ is linked to the running coupling by 
\begin{equation}\label{eq:eta0}
\eta_0=-\partial_t \log \alpha_s(k_{\rm phys}^2)\,.
\end{equation}
At its root \eq{eq:preflowV} is an equation for the dimensionless
effective potential $\hat V = \beta^4 V_k$ in terms of $\hat V_\bot=
\beta^4 V_{\bot,k}$ and $\hat \Delta V=\beta^4 \Delta V_k$. The
infrared RG-scale $k$ naturally turns into the modified RG-scale $\hat
k = k \beta$, that is all scales are measured in temperature units.
Then the flow equation is of the form
\begin{eqnarray}\label{eq:inter}
  \partial_{\hat k} \Delta \hat{ V} = \frac{2 }{3 (2 \pi)^2} \frac{
   (1+\eta_0/5) \hat k^2 }{1+\frac{ g_{k}^2 }{ \hat k^2 } 
    \partial^2_\varphi ( \hat{V}_{\bot} + \Delta \hat{ V})}\,.
\end{eqnarray}
The potential $\hat V$ and hence $\hat\Delta V$ has a
field-independent contribution which is related to the pressure.  For
the present purpose it is irrelevant and we can conveniently normalise
the flow \eq{eq:inter} such that it vanishes at fields where
$\partial_\varphi^2(\hat{V}_{\bot} + \Delta \hat{ V})=0$.  This is
achieved by subtracting $2(1+\eta_0/5)\, \hat k^2 / (3 (2\pi)^2)$ in
\eq{eq:inter} and we are left with
\begin{eqnarray}\label{eq:finalRGeq}
  \partial_{\hat k} \Delta \hat{ V} = -\frac{ 1 }{ 6 \pi^2} 
  \left(1+\frac{\eta_0}{5}\right) \frac{
    \ g_{k}^2 \ 
    \partial^2_\varphi \, ( \hat{V}_{\bot} + \Delta \hat{ V}) }{1
    +\frac{ g_{k}^2 }{ \hat k^2 } 
\partial^2_\varphi \, (\hat{V}_{\bot} + \Delta \hat{ V})}\,,  
\end{eqnarray} 
where we have kept the notation $\partial_{\hat k}\Delta \hat V$ for
$\partial_{\hat k}\Delta \hat V-2(1+\eta_0/5)\, \hat k^2 / (3 (2\pi)^2)$.
In this form it is evident, that the flow vanishes for fields where
$\partial_\varphi^2(\hat{V}_{\bot} + \Delta \hat{ V})=0$, i.e. once a
region of the potential becomes convex, this part is frozen, unless
the external input $\hat V_{\bot}$ triggers the flow again.

We close this section with a discussion
of the qualitative features of \eq{eq:finalRGeq}. It resembles the
flow equation of a real scalar field theory, and due to $V_\bot$, the
flow is initialised in the broken phase. It relies on two external
inputs, $V_\bot$ and $\alpha_s$.

The first input, $\hat V_\bot$, is computed in a perturbative
approximation to the spatial gluon sector, and its computation is
deferred to Appendix~\ref{app:intoutAI}. It is shown in
Fig.~\ref{fig:VPreWeiss3D} for various values of the RG time $\hat k$, 
and approaches the perturbative Weiss potential \cite{Weiss:1980rj} 
for vanishing cutoff $\hat k=0$. 
We have argued that within Polyakov gauge this approximation should
capture the qualitative feature of its contribution to the
Polyakov loop potential. We emphasise again that this is not so for
the question of spatial confinement, and the related potential of the
spatial Wilson loops.

\begin{figure}[h]
 \includegraphics[width=8cm]{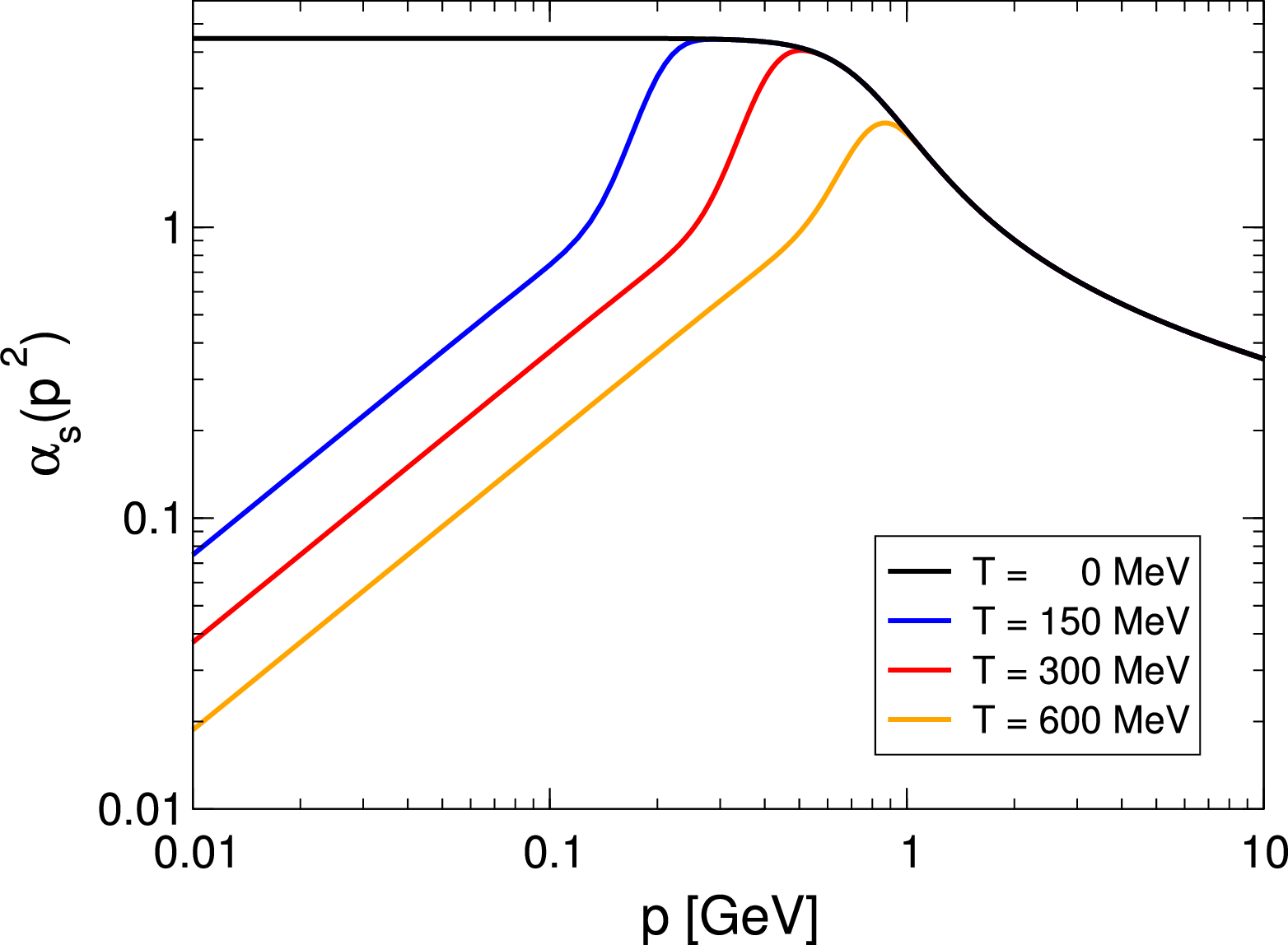}\\
  \caption{$\alpha_s$ for temperatures $T=0,150,300,600$ MeV}
  \label{fig:alpha}
\end{figure}

The second input is $\alpha_s= g_k^2/(4 \pi)$, the running gauge
coupling. It runs with the physical cut-off scale $k_{\rm phys}$
derived in Appendix~\ref{app:match}, $\alpha_s= \alpha_s(k_{\rm
  phys}^2)$. In the present work we model $\alpha_s$ with a
temperature-dependent coupling that runs into a three-dimensional
fixed point $\alpha_{*,3d}k_{\rm phys}/T$ for low cut-off scales
$k_{\rm phys}/T\ll 1$. This choice carries some uncertainty as the
running coupling in Yang-Mills theory is not universal beyond two loop
order. Here we have chosen the Landau gauge couplings $\alpha_{{\rm
    Landau},4d} (k_{\rm phys}^2)$ at cut-off scales $k_{\rm phys}/T\gg
1$, see \cite{Alkofer:2000wg,jan,Fischer:2008uz,von
  Smekal:1997is,Bonnet:2001uh,Lerche:2002ep}. The corresponding
three-dimensional fixed point $\alpha_{*,3d}= 1.12$ is obtained from
\cite{Lerche:2002ep}. A specific choice for such a running coupling is
given in Fig.~\ref{fig:alpha}.  \vspace{.2cm}

\noindent The normalisation of the momentum scale has been done by the
comparison of continuum Landau gauge propagators to their lattice
analogues. Fixing the lattice string tension to $\sqrt{\sigma}=440$
MeV, we are led to the above momentum scales. For a comparison with
the Landau gauge results obtained in \cite{Braun:2007bx} we have also
computed the temperature-dependence of the Polyakov loop by using
$\alpha_{{\rm Landau},4d}$ for all cut-off scales. Indeed, this
over-estimates the strength of $\alpha_s$, as can be seen from
Fig.~\ref{fig:alpha}, However, qualitatively this does not make a
difference: for infrared scales far below the temperature scale, $\hat
k\to 0$, the flow switches off for fields $\varphi$ with
$\partial^2_\varphi(\hat{V}_{\bot} + \Delta \hat{ V})\geq 0$, that is
for the convex part of the potential. This happens both for $g_k^2\to
{\rm const}$, and for $g_k^2(\hat k^2\to 0) \sim \hat k$. In other
words, the minimum of the potential freezes out in this regime. For
the non-convex part of the potential,
$\partial^2_\varphi(\hat{V}_{\bot} + \Delta \hat{ V})<0$, the flow
does not tend to zero but simply flattens the potential, thus
arranging for convexity of the effective potential $V_{\rm
  eff}=V_{k=0}$. The uncertainty in $g^2_k$ is taken into account by
evaluating the limiting cases. Together with the error estimate on the
physical cutoff scale $k_{\rm phys}$ in Appendix~\ref{app:match} this
leads to an estimate for the systematic error of the results presented
below. This error includes that related to our specific choice of the
running coupling. For example, a viable alternative choice to
Fig.~\ref{fig:alpha} is provided by the background field coupling
derived in \cite{Braun:2005uj} which is covered by the above error
estimate.

\section{Integration}\label{sec:integration}

The numerical solution of \eq{eq:finalRGeq} is done on a suitably
chosen grid or parameterisation of $\Delta \hat V$ and its
derivatives. As $\hat V$, $\hat V_\bot$ and $\Delta \hat V$ are
periodic, one is tempted to solve \eq{eq:finalRGeq} in a Fourier
decomposition, see e.g.\ \cite{Braun:2005cn}. However, as can be seen
already at the example of the perturbative Weiss potential $V_W =
V_{\bot,0}$, \eq{eq:Vweiss}, this periodicity is deceiving. The Weiss
potential is polynomial of order four in $\tilde \varphi=\varphi\mod 2
\pi$, its periodicity comes from the periodic $\tilde
\varphi(\varphi)$, \cite{Weiss:1980rj}.  Consequently the third
derivative $\partial_\varphi^3 V_W$ jumps at $\varphi =2 \pi n$ with
$n\in \Z$.  Moreover, $\partial_\varphi^3 V_W[\varphi\to
0_+]=-\partial_\varphi^3 V_W[\varphi\to 0_-]\neq 0$. A periodic
expansion of $V_W$, e.g.\ in trigonometric functions cannot capture
this property at finite order.  This does not only destabilise the
parameterisation, but also fails to capture important physics: the
flow of the position of the minima is proportional to
$\partial_\varphi^3 \hat V$.  This follows from $\partial_t \hat
V[\varphi_{\rm min,k}]=0$. Expanding this identity leads to
\begin{equation}\label{eq:phimin}
  \partial_t \varphi_{min,k} = -\left.\0{\partial_t 
      \hat V'[\varphi]}{\hat V''[\varphi]}
  \right|_{\varphi=\varphi_{\min,k}}\,, 
\end{equation}
where $\hat V'=\partial_\varphi \hat V$ and 
$\hat V''=\partial_\varphi^2 \hat V$. The flow $\partial_t \hat
V'[\varphi]$ is proportional to $\partial_\varphi^3 \hat V$, which
e.g.\ can be seen by taking the $\varphi$-derivative of \eq{eq:flow}.
Hence, as a Fourier-decomposition enforces $\partial_\varphi^3 \hat
V=0$ at any finite order, the minimum does not flow in such an
approximation, and the theory always remains in the deconfined phase.
Note also that the resulting effective potential at $\hat k=0$ for
smooth periodic potentials and flows vanishes identically as it has to
be convex. In the present case this is not so, as the potential is
rather polynomial (in $\tilde \varphi$) and convexity does not enforce
a vanishing effective potential.

In turn, a standard polynomial expansion about the minimum
$\rho_{\min,k}$ already captures the flow towards the confining phase.
Here, however, we use a grid evaluation of the flow of $\Delta
\hat V$ with $\varphi\in [0\,,\,2 \pi]$ while taking special care of the
boundary conditions at $\varphi=0,2 \pi$: we have extrapolated the second
derivative to $\varphi=0$ and $\varphi = 2 \pi$. It suffices to use a first
order extrapolation, and we have explicitly checked that the resulting
flow is insensitive to the precision of the extrapolation.

An alternative procedure is an expansion in terms of Chebyshev
polynomials that also works quite well and is also a very fast and
efficient way of integrating the flow. A comparison between the
results obtained on a grid and with Chebyshev polynomials shows that
both parameterisations agree nicely and the corresponding flows
deviate from each other only for small values of $k$. This is due to
an expected failure of the standard Chebyshev-expansion for those
small $\hat k$ where the position of the minimum is almost settled and
the potential flattens out in the regions that are not convex. This is
better resolved with a grid than with polynomials. On a grid
implementation we see the potential becoming convex as $\hat k \to 0$.

\section{Results}\label{sec:results}

\begin{figure}[h]
  \includegraphics[width=8cm]{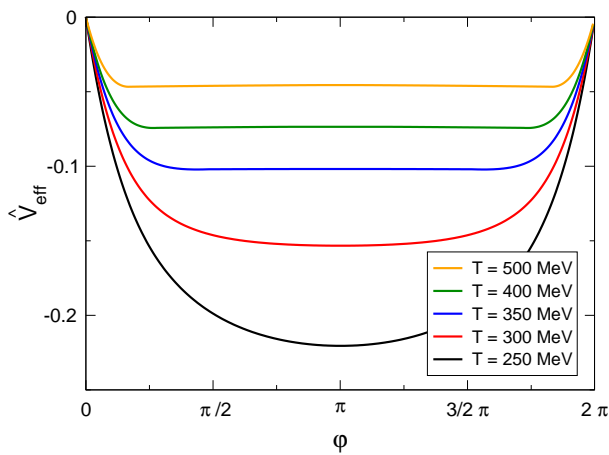}\\
  \caption{Full effective potential $\hat V_{\rm eff}$, 
  normalised to 0 at $\varphi =0$}
  \label{fig:Veff}
\end{figure}

In Fig.~\ref{fig:Veff} we show the full effective potential for
temperatures ranging from $T=500$ MeV in the deconfined phase to
$T=250$ MeV in the confined phase. The expectation value $\langle
\varphi\rangle$ in the center-broken deconfined phase is given by the
transition point between decreasing part of the potential for small
$\varphi$ and the flat region in the middle of the plot. It can also
be explicitly computed from \eq{eq:phimin}. In the center-symmetric
confined phase it is just given by the minimum at $\varphi=\pi$. \\

\begin{figure}[h]
  \includegraphics[width=8cm]{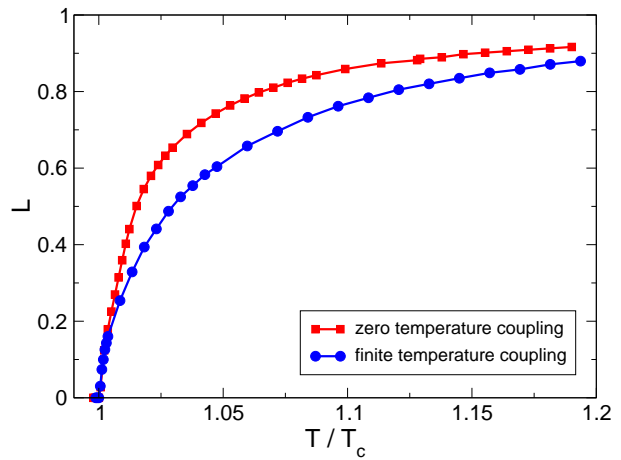}\\
  \caption{Temperature dependence of the Polyakov loop 
$L[\langle A_0\rangle]=\cos(\langle
    \varphi \rangle / 2)$ in $SU(2)$}
  \label{fig:LofT}
\end{figure}

The temperature-dependence of the order parameter $L[\langle
A_0\rangle ] = \cos(\langle \varphi / 2 \rangle)$ is shown in Fig.
\ref{fig:LofT}, and we observe a second order phase transition from
the confined to the deconfined phase at a critical temperature 
\begin{equation} \label{eq:Tc}
T_c =
305^{+ 40}_{-55}\, {\rm MeV},\qquad \quad {T_c}/{\sqrt{\sigma}}
=0.69^{+.04}_{-.12}\,,
\end{equation}
with the string tension $\sqrt{\sigma}=440$ MeV. The
corresponding value on the lattice is ${T_c}/{\sqrt{\sigma}}=.709$,
\cite{Fingberg:1992ju}, and agrees within the errors with our result.
The estimate of the systematic error in \eq{eq:Tc} is dominated by
that of the uncertainty of the identification of $k_{\rm phys}$, see
Appendix~\ref{app:match}.

We would also like to comment on the difference of the
temperature-dependence of $L[\langle A_0\rangle ]$ depicted in
Fig.~\ref{fig:LofT} and that of the Polyakov loop $\langle
L[A_0]\rangle$. It has been shown in section~\ref{sec:QCDinPol} that
in the confined phase they both vanish and both are non-zero in the
deconfined phase. However, the Jensen inequality \eq{eq:jensen}
entails that the present observable $L[\langle A_0\rangle ]$ takes
bigger values than the Polyakov loop $\langle L[A_0]\rangle$, which is 
in agreement with lattice results. 

\begin{figure}[h]
%\vspace{.5cm}
  \includegraphics[width=8cm]{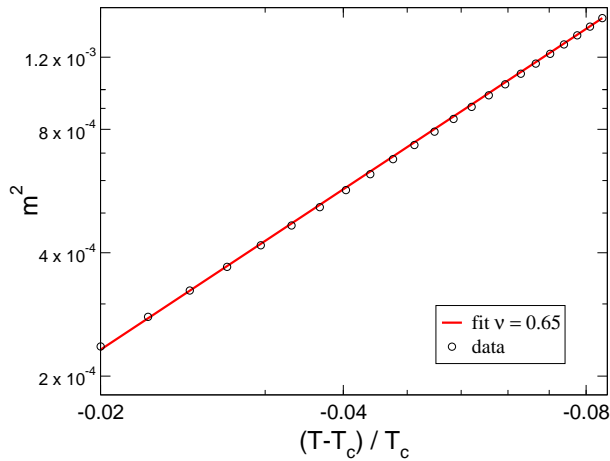}\\
  \caption{Critical exponent $\nu$ from
    $m^2=V''(\varphi_{min,0}) / 2$}
  \label{fig:critexp}
\end{figure}

The critical physics should not depend on this issue. Here we compute
the critical exponent $\nu$, a quantity well-studied in the $O(1)$
model which is in the same universality class as $SU(2)$ Yang-Mills
theory.  Moreover, in Polyakov gauge the effective action
$\Gamma[A_0]$ after integrating-out the spatial gauge field is close
to that of an $O(1)$-model. Studies using the FRG in local potential
approximation with an optimised cut-off for the $O(1)$ model yield
$\nu = 0.65$, see \cite{Litim:2001hk}. The critical exponent is
related to the screening mass of temporal gauge field by
\begin{equation}
m^2(T) \propto |T-T_c|^{2 \nu},
\label{eq:critexp}
\end{equation}
where $m^2 = V''(\varphi_{min,0}) / 2$. We have computed the
temperature-dependence of the screening mass in the confined phase
near the phase transition, and extracted the critical exponent $\nu$
from a linear fit to the data. This is shown in
Fig.~\ref{fig:critexp}. The fit yields the anticipated value of
\begin{equation}
  \nu = 
  0.65^{+0.02}_{-0.01}\,,
\end{equation}
for the critical exponent $\nu$. The critical exponent $\beta$ agrees
within the errors with the Ising exponent
$\beta=0.33$. 

Finally we would like to compare the results obtained here with the
results of \cite{Braun:2007bx}. There, the effective potential $V_{\rm
  eff}[A_0]$ was computed from the flow \cite{jan,Fischer:2008uz} of
Landau gauge propagators \cite{Lerche:2002ep,von
  Smekal:1997is,Bonnet:2001uh} within a background field approach in
Landau-DeWitt gauge.  In this gauge the confining properties of the
theory are encoded in the non-trivial momentum dependence of the gluon
and ghost propagators.  Indeed, in \cite{Braun:2007bx} the effective
potential $V_k$ was computed solely from this momentum dependence and
was not fed back into the flow. In $SU(2)$ Landau gauge Yang-Mills
this is expected to be a good approximation with the exception of
temperatures close to the phase transition, see
\cite{Braun:2007bx}. The back-reaction of the effective potential is
particularly important for the critical physics, and the value of the
critical temperature \cite{BGMP}.

\begin{figure}[h]
  \includegraphics[width=8cm]{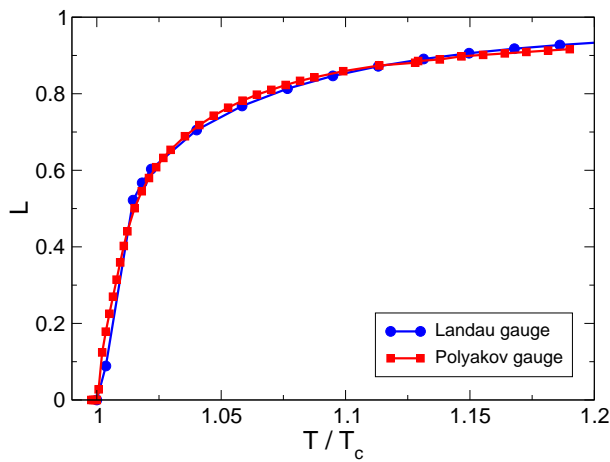}\\
  \caption{Comparison of $L[\langle A_0\rangle]$ computed in Polyakov
    gauge and in Landau-DeWitt gauge from \cite{Braun:2007bx}.}
  \label{fig:compare}
\end{figure}

For the comparison we have computed the present flow with the
zero-temperature running coupling in Fig.~\ref{fig:alpha} for all
temperatures. This mimics the approximation used in
\cite{Braun:2007bx}, which implicitly relies on the zero-temperature
running coupling $\alpha_s$. We also remark that the quantity
$L[\langle A_0\rangle]$ in general is gauge-dependent, and only the
critical temperature derived from it is not. However, in Landau-DeWitt
gauge with backgrounds $A_0$ in Polyakov gauge temporal fluctuations
about this background include those in Polyakov gauge.  For this
reason we might expect a rather quantitative agreement for the
quantity $L[\langle A_0\rangle]$ in both approaches. The results for
the temperature dependence of the Polyakov loop are depicted in
Fig.~\ref{fig:compare}.
The coincidence between the two gauges is very remarkable,
particularly since the mechanisms driving confinement are quite
different in the different approaches, as are the approximations used
in both cases. This provides further support for the respective
results. It also sustains the argument concerning the lack of gauge
dependence made above. The quantitative deviations in the vicinity of
the phase transition are due to the truncation used in
\cite{Braun:2007bx}, that cannot encode the correct critical physics yet, 
as has been already discussed there.

\section{Summary and outlook}\label{sec:summ} 
In the present work we have put forward a formulation of QCD in
Polyakov gauge. We have argued that this gauge is specifically
well-adapted for the investigation of the confinement-deconfinement
phase transition as the order parameter, the Polyakov loop expectation
value $\langle L[A_0]\rangle $, has a simple representation in terms
of the temporal gauge field. Moreover, we have shown that $L[\langle
A_0\rangle]$ also serves as an order parameter. In summary this allows
us to access the phase transition within a simple approximation to the
full effective action.

The computation was done for the gauge group $SU(2)$, where we observe
a second order phase transition at a critical temperature of
$T_c=305^{+40}_{-55}$ MeV, as well as the Ising critical exponents $\nu$
and $\beta$ to the precision achieved within our approximation. The
temperature-dependence of the order parameter $L[\langle A_0\rangle]$
agrees well with a recent computation in Landau gauge
\cite{Braun:2007bx}. This is very remarkable: firstly the latter
computation is technically different as in Landau gauge the full
momentum-dependence of the propagators is needed to cover confinement.
Secondly the order parameter $L[\langle A_0\rangle]$ is gauge
dependent, only the critical temperature is not.

In the present analysis we used several external inputs which we plan
to remove in future work. First of all we proceed with computing the
running coupling within Polyakov gauge, that is the
momentum-dependence of the temporal gauge field. As it is one of the
advantages of the computation in Polyakov gauge that the momentum
dependence of Green functions is rather mild we expect only minor
deviations from the computations shown here. As argued in the present
work, the momentum-dependence of the $A_0$-propagator also gives
access to the string tension. For a description of spatial confinement
one has to treat the spatial components of the gauge field beyond the
present perturbative approximation. Moreover, the present analysis is
extended to $SU(3)$, which is conceptually straightforward but
technically more challenging. For the matter sector one can revert to
the plethora of results with the present renormalisation group
methods, ranging from results in effective theories to that in
QCD-based approaches, see e.g.\
\cite{Berges:2000ew,Pawlowski:1996ch,Braun:2005uj,Braun:2008pi,%
Schaefer:2006sr}.\\[-1ex]

{\em Acknowledgements --} We thank J.~Braun, H.~Gies, F.~Lamprecht,
D.~F.~Litim, A.~Maas and B.-J.~Schaefer for discussions. We thank
O.~Jahn for discussions and collaboration at an early state of this
project. FM acknowledges financial support from the state of
Baden-W\"urttemberg and the Heidelberg Graduate School of Fundamental
Physics.

\begin{appendix}
\section{Faddeev-Popov determinant}
\label{app:FPdet}
From the gauge fixing functionals \eq{eq:Pol1} and \eq{eq:Pol2} we can
compute the Faddeev-Popov determinant given by
\begin{eqnarray}
  \Delta_{FP}[A] = \mathrm{det}\left[\frac{ 
      \delta F^a(A^\omega) }{ \delta \omega^b } \right]\,,
\end{eqnarray}
where $A^\omega$ is the gauge transformed gauge field $A$. For
infinitesimal gauge transformations it is given by
\begin{eqnarray}
  A_\mu^\omega &=& A_\mu - (\partial_\mu \sigma^a 
  + i g A_\mu^b [\sigma^a, \sigma^b ]) \omega^a\,.
\end{eqnarray}
In the following we use the representation $\omega^a \sigma^a =
\omega^+ \sigma^- + \omega^- \sigma^+ + \omega^3 \sigma^3 $, and the 
related derivatives w.r.t.\ $\omega^{\pm},\omega^3$. 
The matrix elements related to $\omega$-derivatives of $F^+$ read 
\begin{eqnarray}\nonumber 
  \frac{\delta F^+ (A^\omega)}{ \delta \omega^+ } 
  &=& - \Tr\,\sigma^+ \left( 
    \partial_0 \sigma^- + i A_0^3 [\sigma^-, \sigma^3] 
  \right) \,,\\\nonumber 
  \frac{\delta F^+ (A^\omega)}{ \delta \omega^- }
&=&0\,,\\
  \frac{\delta F^+ (A^\omega)}{ \delta \omega^3 } 
  &=& -\Tr\,\sigma^+ \left( \partial_0 
    \sigma^3 + i A_0^+ [\sigma^3, \sigma^-] \right)\,.  \hspace{.5cm}
\label{eq:coef+}\end{eqnarray}
Analogously we get for the $\omega$-derivatives of $F^-$ 
\begin{eqnarray}\nonumber
  \frac{\delta F^- (A^\omega)}{ \delta \omega^+ } 
&=&0\,,\\\nonumber
  \frac{\delta F^- (A^\omega)}{ \delta \omega^- } 
&=&  -\Tr\,\sigma^- \left( \partial_0 \sigma^+ 
    + i A_0^3 [\sigma^+, \sigma^3] \right)\,,\\
  \frac{\delta F^- (A^\omega)}{ \delta \omega^3 } 
&=& -\Tr\,\sigma^- \left( \partial_0 
    \sigma^3 + i A_0^- [\sigma^3, \sigma^+] \right)\,.
\label{eq:coef-}\end{eqnarray}
The $\omega$-derivatives of $F^3$ yield long
expressions, and we only display the parts proportional to
$\partial_0 \Tr \,\sigma^3 A_0$, where we have abbreviated additional terms
proportional to the spatial gauge fields by dots, 
\begin{eqnarray}\nonumber  
  \frac{\delta F^3 (A^\omega)}{ \delta \omega^+ } 
&=&
  -i \partial_0 A_0^- \,
  \Tr \, \sigma^3  [\sigma^-, \sigma^+] + \cdots\,,  \\\nonumber
  \frac{\delta F^3 (A^\omega)}{ \delta \omega^- } 
&=&  - i \partial_0  A_0^+ \,\Tr \, 
  \sigma^3  [\sigma^+, \sigma^-]  + \cdots\,, \\
  \frac{\delta F^3 (A^\omega)}{ \delta \omega^3 }
&=&-	2 \partial_0^2 + \cdots \,.
\label{eq:coef3}\end{eqnarray}
\begin{widetext}
Evaluating the traces \eq{eq:coef+},\eq{eq:coef-},\eq{eq:coef3}
we can compute the Faddeev-Popov determinant. 
Again we only concentrate
on the terms dependent on $A_0$, and use the gauge fixing condition
$A_0^+ = A_0^- = 0$ for eliminating some  of the off-diagonal elements, 
\begin{eqnarray}
  \Delta_{FP}[A]  &=& -
  \mathrm{det}\left[ \left(
  \begin{array}{ccc}
    \partial_0 +ig A_0^3 & 0 & 0\\
    0 & \partial_0 - ig A_0^3 & 0\\
    -4 ig \int dx_0 \partial_1 A_1^- + \cdots & 
    4 i g \int dx_0 \partial_1 A_1^+ + \cdots & 1/2( \partial_0^2 
    + \int dx_0 \partial_1^2 + \cdots)
  \end{array}
\right) \right] \nonumber\\
&=& - \mathrm{det}[(\partial_0 + ig A_0^3) 
(\partial_0 -ig A_0^3)
\frac{1}{2} \left( \partial_0^2 + \int dx_0 \partial_1^2 
  + \int dx_0 dx_1 \partial_2^2+ \int dx_0 dx_1 dx_2 
\partial_3^2 \right)]
\end{eqnarray}
\end{widetext}
Using the third gauge fixing condition, $\partial_{0} A_{0}^3 = 0$, we can write the Faddeev-Popov
determinant as
\begin{eqnarray}\nonumber 
  \Delta_{FP}[A]  &=&
      \frac{1}{2}\mathrm{det} \left[ \left(
        \partial_0^2 + \left(g A_0^3\right)^2\right)\right]
    \mathrm{det}[\left( \partial_0^2 + \cdots \right)]\,.
\label{eq:FPdet0} \end{eqnarray}
We note that the second determinant in \eq{eq:FPdet0} is independent of
the gauge fields and hence can be absorbed in the normalisation of the
path integral. The first determinant is evaluated in frequency space, 
we get 
\begin{equation}
\prod_{\vec x}   \left( (g A_0^3 (\vec x))  \prod_{n=1}^{n 
      = \infty} \left((2 \pi T n)^2 - (g A_0^3 (\vec x))^2\right) 
  \right)^2\,. 
\label{eq:FPdet1}\end{equation}
Multiplying the determinant \eq{eq:FPdet1} with a further constant
normalisation 
\begin{eqnarray}
  \mathcal{N} = \left( \prod_{n=1}^{n = 
      \infty} (2 \pi T n)^2 \right)^{-2}\,,
\end{eqnarray}
we arrive at 
\begin{equation}
  \mathcal{N} \mathrm{det} \left[ G_{A_0}\right] =
  \prod_{x}  (g A_0^3 (x))^2 \prod\limits_{n=1}^{n 
    = \infty} \left(1 - \left( \frac{g A_0^3 (x)}{ 2 \pi n T} 
    \right)^2 \right)\,.
\label{eq:FPdet2}\end{equation}
\Eq{eq:FPdet2} is just a product representation of the sine-function,
$\sin(x) = x \prod_{n=1}^{n = \infty} \left(1 - \frac{x^2}{ (\pi n)^2
  }\right)$, and the final result for the Faddeev-Popov determinant is
\begin{eqnarray} 
  \Delta_{FP}[A] = \mathcal{N}' (2 T)^2 \left[ \prod_{x} 
    \sin^2 \left( \frac{g A_0^3 (x)}{ 2 T } \right) \right],
\end{eqnarray}
where $\mathcal{N}' $ is a further normalisation constant. 

 \begin{widetext} 
\section{Integrating out spatial gluons}
\label{app:intoutAI}
After integrating out the longitudinal gauge fields the action 
$S_{\rm eff}= \frac{1}{4}\int_T d^4x 
  F_{\perp, \mu \nu}^a
  F_{\perp,\mu \nu}^a$ reads 
\begin{equation} 
  S_{\rm eff}=  - \frac{1}{2} 
  \beta\int  d^3 x\, Z_0 A_0 \vec \partial^2 A_0
  - \frac{1}{2} \int_T d^4x\,  
  A_i^a \left[ (\partial_0^2 + \vec \partial^2) \delta_{ij} -
  \partial_i \partial_j + 
  2 g f^{a3b} (A_0 \partial_0 +g^2 A_0^2 (\delta^{ab} - 
  \delta^{a3}\delta^{b3})\delta_{ij}\right] A_j^b +O(A_i^3)  
\end{equation} 
\end{widetext}
Writing $A_0^3 = \varphi/(g \beta ) + a_0$, where $\varphi$ is a constant and
$a_0$ the fluctuating field, this expression is given to second order
in the fluctuating fields by
\begin{eqnarray} S_{YM} &\approx& \frac{1}{2} \int d\tau d^3x\, 
\left\{ Z_0 (\vec
  \partial a_0)^2 -2 \varphi f^{a3c} (\partial_0 A_i^a) A_i^c+ \right.
  \nonumber\\
  & & \hspace{1.9cm} \varphi^2 (\delta^{ab} - 
\delta^{a3}\delta^{b3})A_i^a A_i^b -
  \nonumber \\
  & & \hspace{1.9cm} \left. A_i^a \left( (\partial_0^2 + \vec \partial^2) 
\delta_{ij} -
\partial_i \partial_j \right)A_j^a \right\} \nonumber \\
&=& \frac{1}{2} \int d\tau d^3x \ \left\{ (\vec \partial a_0)^2 -
  A_i^a ( \vec \partial^2 - \partial_i \partial_j) A_j^a - \right.
\nonumber\\
& & \hspace{1.9cm}\left. A_i^a D_0^{ac}D_0^{cb} A_i^b \right\}, 
\end{eqnarray} 
where we have defined
\begin{eqnarray} 
  D_0^{ab} = \partial_0 \delta^{ab} + A^3_0 g f^{a3b} .
\end{eqnarray} 
In the present work we neglect back-reactions of the $A_0$ potential
on the transversal gauge fields. Assuming an expansion around $A_i^a =
0$, $\Gamma^{(2)}$ is block-diagonal, like the regulators, cf. eq.
(\ref{eq:cutoffs}), and we can decompose the flow equation
(\ref{eq:flow}) into a sum of two contributions, schematically written
as
\begin{eqnarray}\label{eq:FRG_before_intout}
  \partial_t{\Gamma}_{k} &=&
  \frac{1}{2}\mathrm{Tr} \left(\frac{1}{\Gamma_{k}^{(2)} + R_A}
  \right)_{00}
  \partial_t R_k + \nonumber \\
  & &\hspace{2cm}\Tr\, \partial_t \left[\ln (S_{YM}^{(2)} + R_{A}) 
  \right]_{ii}.  
\end{eqnarray} 
The first term on the rhs encodes the quantum fluctuations of $A_0$,
the second line encodes those of the transversal spatial components of
the gauge field. In the present truncation the second line is a total
derivative w.r.t. $t$, and does not receive contributions from the
first term. Therefore we can evaluate the flow of the second
contribution, and use its output $V_{\bot,k}(A_0)$ as an input for the
remaining flow.

The computation is done for the regulators \eq{eq:cutoffs}.  As
explained below \eq{eq:stringmomentum} in section~\ref{sec:approx},
the cut-off parameters $k$, and $k_\bot $ in $R_k$ for the
fluctuations of $A_0$ and $R_{k,\bot}$ for the fluctuations of $\vec
A_\bot$ respectively satisfy a non-trivial relation $k_\bot =k_\bot
(k)$ for coinciding physical infrared cut-offs $k_0$ for $A_0$ and
$k_\bot $ for $\vec A_\bot$. The computation is similar to those done
in one loop perturbation theory in $SU(2)$ by Weiss
\cite{Weiss:1980rj}, the only difference being the infrared cut-off.
We infer from the second line in \eq{eq:FRG_before_intout} that
\begin{eqnarray}\label{eq:preVbot}  
  V_{\bot,k} &=& V_{\bot,\Lambda_{\rm UV}}
  +\left. 
    \012 \Tr\, \left[\ln (S_{YM}^{(2)} + R_{A})\right]_{ii}
  \right|^k_{\Lambda_{\rm UV}} 
  \\\nonumber  
  &=& V_W + T \sum_{n} \int \frac{d^3p}{(2\pi)^3} \theta (k_\bot ^2 -
  \vec p^2) \ln ( k_\bot ^2 + D_0^2)\,.
\end{eqnarray}
In \eq{eq:preVbot} we have used that $V_{\bot,\Lambda_{\rm UV}\to
  \infty}=0$ up to a constant term, and have added and subtracted the
Weiss potential $V_W$ \cite{Weiss:1980rj},
\begin{equation}\label{eq:Vweiss}
  V_W(\varphi) = -(\tilde\varphi-\pi)^2 / (6 \beta^4) +
  (\tilde\varphi-\pi)^4 / (12 \pi^2 \beta^4)\,,  
\end{equation}
with the dimensionless $\varphi=g\beta A_0$, and $\tilde
\varphi=\varphi \mod 2 \pi$.  Alternatively one can simple put
$\Lambda_{\rm UV}=0$, even though this seems to be counter-intuitive.
We also have used that with \eq{eq:transverse} it follows $\tr\,
\Pi_\bot =2$. Performing the Matsubara sum and neglecting terms
independent of the temporal gauge fields, the resulting effective
potential is given by
\begin{eqnarray}\label{eq:Vbot}
  V_{\bot,k} &=& \frac{4 T}{(2\pi)^2} \int_0^{k_\bot } dp p^2
  \left\{ \mathrm{ln}\Bigl(1-2\cos(\varphi) e^{-\beta k_\bot } 
   \right. \\
  & & \hspace{-.5cm} \left. + e^{-2 \beta
      k_\bot }\Bigr)
    - \mathrm{ln}(1-2\cos(\varphi) e^{-\beta p} + e^{-2 \beta p})
  \right\} + V_W \,.
\nonumber \end{eqnarray} 
From \eq{eq:Vbot} we deduce that the potential $ V_{k_\bot }$
approaches $V_W$ in the limit $k \to 0$ and vanishes like $e^{-\beta
  k_\bot } \cos(\varphi)$ for $k \to \infty$.  From eq.
(\ref{eq:FRG_before_intout}) we can now extract the flow of the
effective potential, by setting $V_{\mathrm{eff},k} = \Delta V_{k} +
V_{\bot,k}$. Then we get
\begin{equation} \label{eq:flowDeltaVapp}
  \partial_t \Delta V_{k}=\012 
  \int \0{d^3 p}{(2 \pi)^3}
  \frac{(\eta_0(k^2-\vec p^2)+ 2k^2 )\theta (k^2 - \vec p^2)
  }{k^2 + g_k^2 \beta^2 (\Delta V_{k}'' + V_{\bot,k}'') }\,,
\end{equation} 
with the input $V_{\bot,k}$ given in \eq{eq:Vbot} and
$\eta_0=\partial_t \ln Z_0$.  The factor $g^2 \beta^2$ arises from the
fact that we parametrise the potential in terms of $\varphi$ rather
than in $A_0$, and $g_k^2=g^2/Z_0$ is nothing but the running coupling
at momentum $\vec p^2\sim k_{\rm phys}^2$. Thus we estimate
$g_k^2=4\pi \alpha_s(\vec p^2=k_{\rm phys}^2)$. Note that $g_k$ is an
RG-invariant. The momentum integration can be performed analytically,
and we are led to
\begin{eqnarray}\label{eq:preflowVapp}
  \beta \partial_k \Delta V_k
  = \frac{2}{3 (2 \pi)^2} \frac{(1+\eta_0/5) k^2 }{1+\frac{ g_{k}^2 
      \beta^2}{ k^2 }
    \partial^2_{\varphi} ( V_{\bot,k} + \Delta V_k)}\,, 
\end{eqnarray}
where $\eta_0$ is given by
\begin{equation}\label{eq:eta0app}
\eta_0=-\partial_t \log \alpha_s\,, 
\end{equation}
as the consistent choice in the given truncation.

\section{Matching scales}
\label{app:match}

\begin{figure}[h]
\vspace{.2cm}
  \includegraphics[width=8cm]{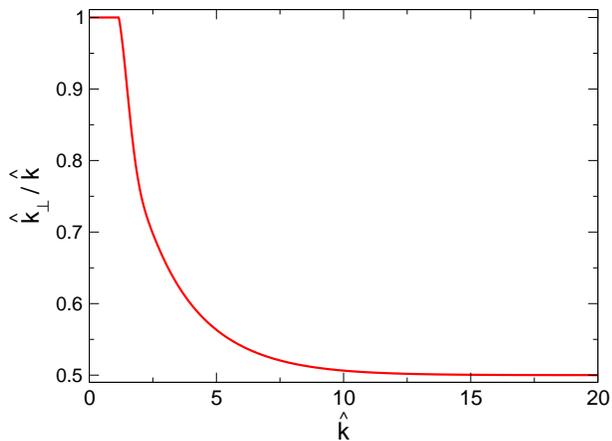}\\
  \caption{${\hat k}_\bot / \hat k$ as function of $\hat k$. }
  \label{fig:kbotk}
\end{figure}

The flow of the temporal component of the gauge field, $A_0(\vec x)$,
is computed with a three-dimensional regulator, see \eq{eq:cutoffs}.
In Polyakov gauge $A_0(\vec x)$ only depends on the spatial
coordinates, whereas the spatial components $A_\bot(x)$ are
four-dimensional fields. For cut-off scales far lower than the
temperature, $k/T\ll 1$, also the spatial gauge fields are effectively
three-dimensional fields as only the Matsubara zero mode propagates.
Hence in this regime we can identify $k=k_\bot$. For large cut-off
scales, $k/T\gg 1$, the $A_0$-flow decouples from the theory.  A
comparison between the two flows can only be done after the summation
of the spatial flow over the Matsubara frequencies.  In the asymptotic
regime $k/T\gg 1$ this leads to the relation
\begin{equation}\label{eq:kTinf}
 \frac{1}{k} \simeq \sum_{n=-\infty}^{\infty} 
  \frac{1}{\omega_n^2 + k_\bot ^2}\to \frac{1}{2 k_\bot}\,,
\end{equation} 
The crossover between these asymptotic regimes happens at about $k/T=
1$. This crossover is implemented with the help of an appropriately
chosen interpolating function $f$,
\begin{eqnarray}\label{eq:compare}
 \frac{T}{k^2} f(k/T) &=& T \sum_{n=-\infty}^{\infty} 
  \frac{1}{\omega_n^2 + k_\bot ^2}\,,
\end{eqnarray}
A natural choice for $f(k/T)$ is depicted in Fig.~\ref{fig:kbotk}, and
has been used in the computation. A more sophisticated adjustment of
the relative scales can be performed within a comparison of the flow
of momentum-dependent observables such as the wave function
renormalisation $Z_0$. The peak of these flows in momentum space is
directly related to the cut-off scale. Indeed, the function $f$
carries the physical information of the peak of the flow at some
momentum scale. Scanning the set of $f$ gives some further access to
the uncertainty in such a procedure.
The effective cut-off scales $k_{\rm phys}(k_0)$ and $k_{\bot ,\rm
  phys}(k_\bot )$ in the flows of the temporal gluons and of spatial
gluons respectively do not match in general. If solving the flow
within a local truncation as chosen in the present work we have to
identify the two effective cut-off scales, $k_{\rm phys}(k_0)=k_{\bot
  ,\rm phys}(k_\bot )=k_{\rm phys}$, leading to a non-trivial relation
$k_0=k_0(k_\bot )$.  Moreover, the effective cut-off scale has to be
used in the running coupling $\alpha_s=\alpha_s(\vec p^2=k_{\rm
  phys}^2)$.

\vspace{0.5cm}
\begin{figure}[h]
  \includegraphics[width=8cm]{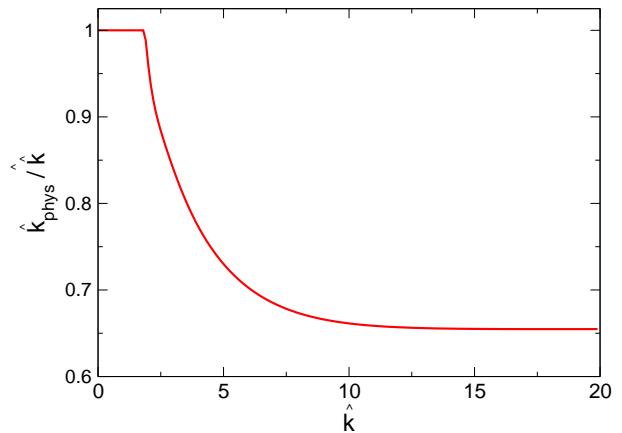}\\
  \caption{$\hat k_{\rm phys}(\hat k )$ 
from the comparison of flows with 
three-dimensional regulators and four-dimensional regulators.}
  \label{fig:kskphys}
\end{figure}

It is left to determine the physical cut-off scale $k_{\rm phys}$ from
either the flow of the spatial gauge fields as $k_{\bot,\rm
  phys}(k_\bot )$ or from the temporal flow $k_{0,\rm phys}(k_0)$.  We
first discuss the spatial flow. For an optimised regulator depending
on all momentum directions, $p^2$, we have the relation $k_{\rm
  phys}=k_\bot $. Hence the relation $k_{\bot,\rm phys}(k_\bot )$ can
be computed if comparing the flows for a specific observable with
three-dimensional regulator $R_{{\rm opt},k_\bot }(\vec p^2)$,
\eq{eq:opt}, with flows with four-dimensional regulator $R_{{\rm
    opt},k_{\rm phys}}(p^2)$. Here, as a model example, we choose the
effective potential of a $\phi^4$-theory. This leads to the relation
$k_{\rm phys}(k_\bot )$ displayed in Fig~\ref{fig:kskphys}. We remark
that the relation in Fig.~\ref{fig:kskphys} depends on the dimension
$d$ of the theory, and flatten to $k_{\rm phys}(k)=k$ for
$d\to\infty$. In other words, $\lim_{k\to \infty} k_{\rm phys}(k)/k$
is proportional to $d/(d-1)$. Moreover, for momentum-dependent
observables the crossover rather resembles the relation $k_\bot(k_0)$
as it is more sensitive to the propagator than to the momentum
integral of the propagator. Indeed, for the three-dimensional field
$A_0(\vec x)$ the cut-off scale $k_0$ is another natural choice for
the physical cut-off scale, $k_{0,\rm phys}(k_0)=k_0$, even though it
underestimates the importance of the spatial flow for the correlations
of the temporal gauge field. In summary, we take the above two
extremal choices $k_{\rm phys}=k_{\bot,\rm phys}$ depicted in
Fig.~\ref{fig:kskphys} and $k_{\rm phys}= k_{0,\rm phys}(k_0)=k_0$ as
a broad estimate of the systematic error in the present computation.

\end{appendix}

\end{document}